\documentclass{vgtc}                          

\graphicspath{{figures/}{pictures/}{images/}{./}} 

\usepackage{times}                     

\usepackage{tabu}                      
\usepackage{booktabs}                  
\usepackage{lipsum}                    
\usepackage{mwe}                       

\usepackage{mathptmx}                  

\usepackage{xcolor}

\onlineid{1016}

\vgtccategory{Research}

\vgtcinsertpkg

\title{When Avatars Have Personality: Effects on Engagement and Communication in Immersive Medical Training}

\author{
Julia S. Dollis\thanks{e-mail: juliadollis@discente.ufg.br}
\and Iago A. Brito\thanks{e-mail: iagoalves@discente.ufg.br}
\and Fernanda B. F\"arber\thanks{e-mail: fernandabufon@gmail.com}
\and Pedro S. F. B. Ribeiro\thanks{e-mail: schindler@discente.ufg.br}
\and Gustavo H. W. Barbosa \thanks{e-mail: guga.webster@gmail.com}
\and Andressa A. Bastos\thanks{e-mail: andressa.bastos08@gmail.com}
\and Rafael T. Sousa\thanks{e-mail: rafaelsousa@ufmt.br}
\and Arlindo R. Galv\~ao Filho\thanks{e-mail: arlindogalvao@ufg.br}
}
\affiliation{\scriptsize Advanced Knowledge Center for Immersive Technologies (AKCIT)\\Federal University of Goiás (UFG)}

\teaser{
  \centering
  \includegraphics[width=\linewidth]{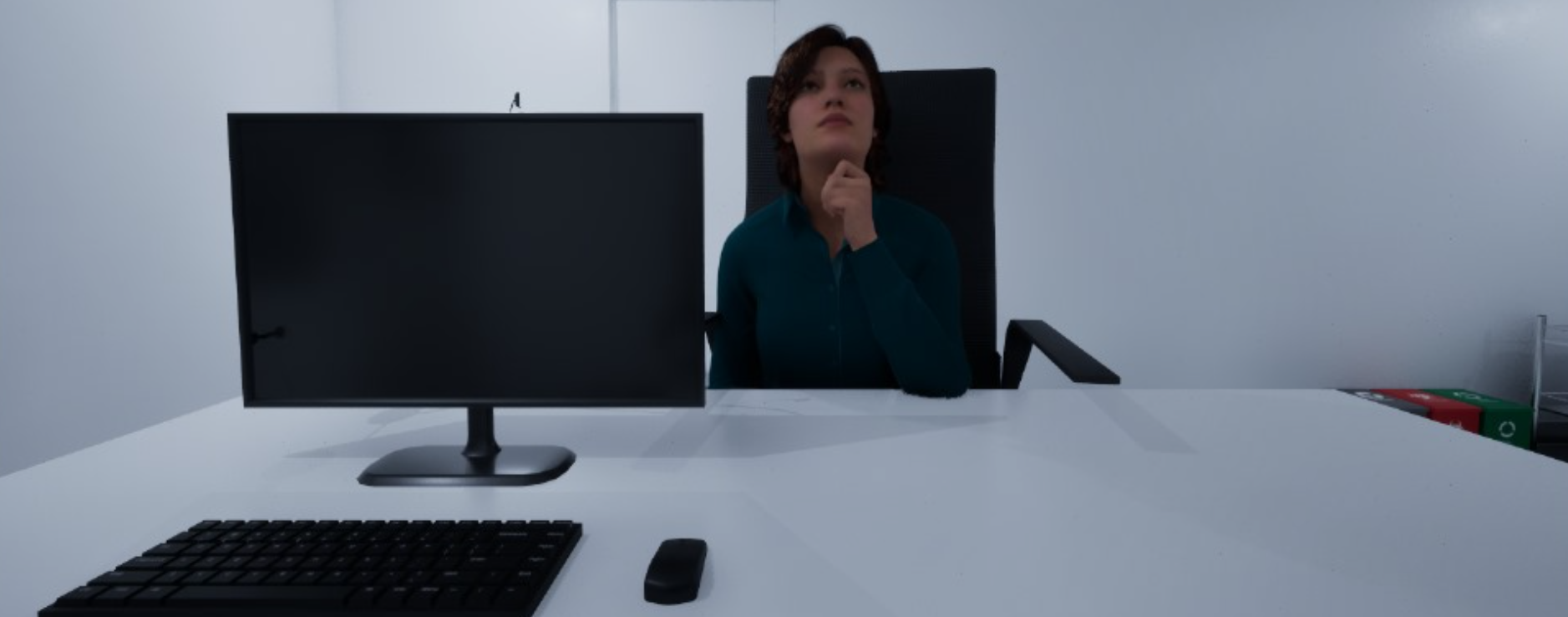}
  \caption{The physician’s perspective in our immersive VR consultation. A large language model generates the virtual patient's dialogue and personality in real-time, moving beyond static, scripted encounters.}
  \label{fig:teaser}
}

\abstract{
    While virtual reality (VR) excels at simulating physical environments, its effectiveness for training complex interpersonal skills is limited by a lack of psychologically plausible virtual humans. This gap is particularly critical in medical education, where communication is a core clinical competency. This paper introduces a framework that integrates large language models (LLMs) into immersive VR to create medically coherent virtual patients with distinct, consistent personalities, based on a modular architecture that decouples personality from clinical data. We evaluated the system in a mixed-methods, within-subjects study with licensed physicians conducting simulated consultations. Results suggest that the approach is feasible and perceived as a rewarding and effective training enhancement. Our analysis highlights key design principles, including a “realism–verbosity paradox” and the importance of challenges being perceived as clinically authentic to support learning.
}


\begin{document}


\firstsection{Introduction}
\maketitle

Technological advances in immersive environments have transformed virtual reality into a powerful platform capable of reproducing highly realistic physical environments and immersive scenarios. By integrating high-fidelity graphics, motion tracking, and responsive interaction design, virtual reality has found successful applications across domains such as industrial training \cite{radhakrishnan2021systematic}, security \cite{10257288} and healthcare \cite{ekstrand2018immersivem}. Its ability to replicate spatial, visual, and procedural aspects of real-world scenarios with high accuracy has made it particularly valuable in contexts that require experiential learning under safe and controlled conditions, positioning virtual reality as a promising solution for skill acquisition, procedural rehearsal, and decision-making practice where realism is essential to effective transfer of learning \cite{gunn2018use}.

While these recent advances in rendering techniques have markedly improved the visual and physical fidelity of VR systems, the psychological, emotional, and social dimensions of digital humans remain comparatively underexplored \cite{vcekic2025virtual}. Realistic textures, animations, and gestures can enhance immersion, yet interactions with digital humans are not defined solely by appearance. Human communication inherently involves complex social, emotional, and psychological processes, among which personality plays a fundamental role in shaping how individuals convey information, respond to others, and co-construct meaning during interaction \cite{hassan2019anatomy, dhillon2023impact}. The absence of mechanisms to represent diverse and consistent personalities risks VR simulations to producing environments that are visually compelling but socially impoverished, thereby limiting their effectiveness in scenarios that rely on authentic interpersonal dynamics.

 This gap becomes particularly critical when VR is used to prepare individuals for real-world sociable interactions. In medical consultations, for instance, doctors must not only identify symptoms and apply clinical reasoning but also adapt their communication strategies to patients who differ widely in personality, emotional expressiveness, openness, or levels of trust. Personality shapes how patients present symptoms, respond to questioning, and engage emotionally during clinical encounters \cite{flynn2007personality}. As a result, simulations that neglect these crucial dimensions fail to replicate the full spectrum of a clinical encounter, limiting their effectiveness in developing the nuanced interpersonal competencies essential for effective patient care.

In parallel, recent advancements in natural language processing with the advent of large language models (LLMs) have opened a promising pathway to simulate human behavior in digital environments \cite{xie2024can}. These models have demonstrated strong capabilities in producing coherent, contextually appropriate, and personality-consistent dialogue \cite{park2023generative}, allowing a move from rigid scripts to dynamic and adaptive communication. Embedding this technology into VR environments has the potential to overcome the current lack of realistic personality in digital humans, unlocking a new level of social faithfulness that is crucial for studying interpersonal interactions.

In this work, we introduce a framework that integrates LLM-driven personality models into immersive VR medical consultations to create more authentic and challenging training scenarios. Our system allows medical professionals to engage with psychologically plausible virtual patients, enabling us to systematically investigate how variations in patient personality-such as cooperativeness, anxiety, or resistance—influence a physician’s diagnostic strategies and perceived realism. We evaluate these variations by capturing physician feedback using a series of Likert-scale questionnaires designed to assess each virtual patient individually and a direct comparison between different personality profiles. Moreover, to address the inherent challenge of collecting a high amount of real doctor–patient interactions, we propose a novel method for synthetic data augmentation. By leveraging LLMs to generate and analyze thousands of simulated consultations, we conduct a large-scale assessment of our personality models' robustness, consistency, and adaptability across diverse clinical scenarios, ensuring both behavioral and medical coherence.

Building on prior evidence that patient personality shapes communication patterns in clinical encounters \cite{redelmeier2021understanding, oldham2020personality, lee2025personality}, we posit the following hypotheses:

\textbf{H1:} It is feasible to simulate patients with different personality profiles in virtual agents using large language models embedded in immersive VR environments, maintaining both behavioral consistency and coherence of disease symptoms.

\textbf{H2:} Virtual patients with distinct personalities will elicit differences in doctors’ interaction strategies, including questioning style, dialogue flow, and engagement levels.

To test these hypotheses, we designed and implemented an immersive VR environment simulating medical consultations, as illustrated in \cref{fig:teaser}, in which virtual patients are powered by LLMs to exhibit consistent and distinct personality profiles (e.g., introverted, extroverted, resistant, cooperative). We conducted a mixed-method evaluation combining quantitative measures of engagement and realism with qualitative feedback from participating doctors.

\textbf{Contribution Statement.} This work makes three main contributions:

(1) We propose a novel framework that integrates LLM-based personality modeling into VR medical consultations, moving beyond scripted or procedural communication.

(2) We provide a controlled experimental evaluation of the role of patient personality in shaping interaction dynamics, perceived realism, and training value in VR medical simulations.

(3) Our pilot study provides preliminary evidence suggesting that personality is a significant factor in designing effective VR training environments, with significant implications for both medical education and human-computer interaction.

\section{Related Works}

\subsection{Virtual Reality in Medical Training}

Virtual reality (VR) has become an established tool in medical education, offering immersive and controlled environments where learners can acquire procedural, diagnostic, and communication skills without risks to real patients \cite{kyaw2019virtual, ekstrand2018immersivem, sattar2019effects, muhling2025personality}. Numerous studies have demonstrated its effectiveness in domains such as surgical training, anatomy learning, and interprofessional collaboration, often reporting improvements in skill acquisition and learner engagement compared to traditional methods \cite{mergen2024reviewing, sung2024effectiveness, muhling2025comparing}. VR-based scenarios have also been used to train delicate competencies, including empathy and breaking bad news, showing that simulated encounters with virtual humans can elicit measurable behavioral changes \cite{guetterman2019medical, carnell2022informing}.

In recent years, artificial intelligence (AI) has been integrated into VR training systems to extend their adaptability and realism. AI-driven virtual patients and tutors can generate dynamic feedback, tailor scenarios to individual learner performance, and support assessment of communication quality \cite{borg2025virtual, kapadia2024evaluation}. For example, LLM-powered agents have been proposed to evaluate clinical decision-making, provide formative feedback, or simulate standardized patients at scale \cite{cook2025virtual, weisman2025development}. Despite these advances, most AI-enhanced VR applications remain focused on procedural or technical training (e.g., orthopedics, emergency protocols), with relatively limited exploration of psychosocial and interpersonal challenges, dimensions that are critical in real-world consultations. This gap motivates the need to examine how AI-driven virtual patients can embody nuanced traits such as personality to enrich social realism in VR medical training.

\subsection{Personality Simulation in Digital Agents}

Parallel to advances in VR, a growing body of work has investigated how digital agents, particularly those powered by large language models, can be endowed with distinct personalities and communication styles. Personality expression has been shown to significantly influence engagement, trust, and learning outcomes in educational agents \cite{sonlu2024effects, zhu2025social}. Recent frameworks such as PATIENTSIM \cite{kyung2025patientsim} and Adaptive-VP \cite{lee2025adaptive} highlight the potential of LLMs to simulate clinically relevant personas, ranging from anxious or impatient patients to individuals with limited recall or language proficiency. These systems move beyond scripted dialogues by capturing variation in emotional expression, conversational tone, and interaction difficulty, thereby creating more authentic and challenging training opportunities. Similarly, other works have modeled “difficult” communication styles, such as the Satir-based accuser or rationalizer, to expose learners to demanding interpersonal dynamics \cite{bodonhelyi2025modeling}.

Outside the medical domain, personality-driven agents have also been employed for training in high-stakes contexts such as firefighting and disaster response, where realistic interpersonal stressors and decision-making under pressure are essential \cite{oliveira2024exploring}. These studies consistently demonstrate that embedding personality traits in digital agents enriches user engagement and promotes transfer of social skills. However, while the impact of personality has been explored in LLM-driven chat-based agents and 2D training systems, its systematic integration into immersive VR medical training remains underdeveloped. This underscores the importance of investigating how patient personality in VR consultations shapes doctors’ strategies, perceptions of realism, and educational outcomes.

\begin{figure*}
    \centering
    \includegraphics[width=1\linewidth]{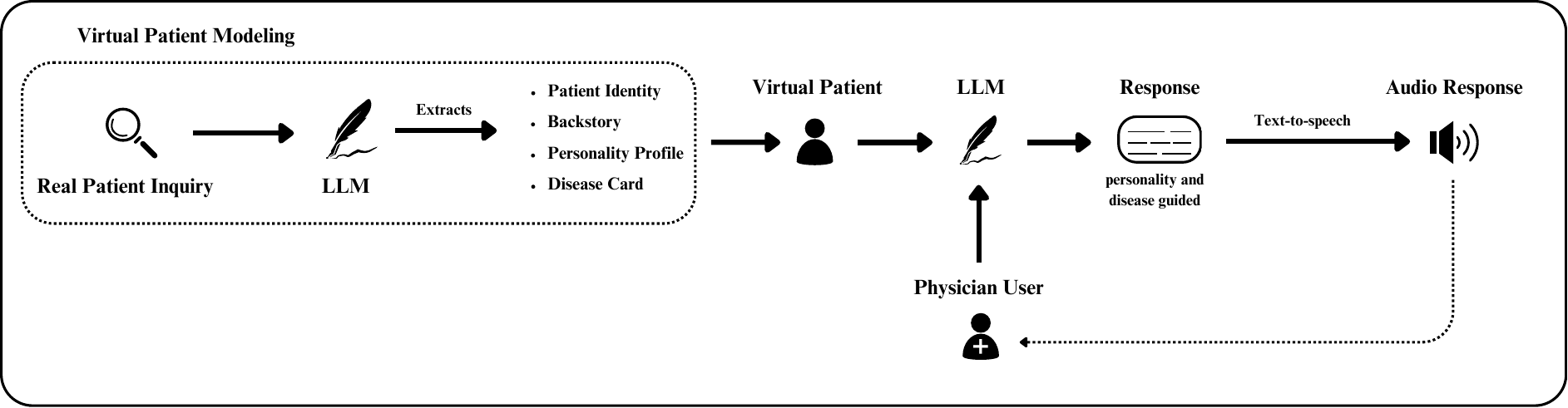}
    \caption{The architecture of our interactive patient simulation pipeline. The Virtual Patient Modeling stage (left) utilizes a real patient inquiry to generate a medically and behaviorally coherent patient. During the Physician Interaction stage (right), a patient response is generated using an LLM that is synthesized into speech via a Text-to-Speech (TTS) module. The corresponding Speech-to-Text (STT) module, which processes the physician's audio input, is omitted for visual clarity.}
    \label{fig:pipeline}
\end{figure*}

\section{Methods}


To test our hypotheses, we designed, implemented, and evaluated a comprehensive VR medical training framework. This section details our methodology. We begin by describing the technical implementation of the simulation, starting with the VR environment. We then detail our core contribution in virtual patient modeling, which outlines the architecture used to generate patients with distinct and consistent backstories, personality profiles, and disease. Following this, we present our experiment design, specifying the clinical scenarios, participant recruitment, within-subjects procedure, and the mixed-method measures used in our user study. Finally, we introduce our novel method for large-scale synthetic data analysis in LLM-driven generated personalities, an approach designed to assess the robustness and scalability of our personality models. The overall framework is presented in \cref{fig:pipeline}.

\subsection{VR Environment} \label{ssec:vr_environment}

To provide a realistic and immersive setting for the simulated consultations, we implemented a high-fidelity virtual consultation environment using Unreal Engine, a widely adopted platform for interactive VR applications across domains such as training, simulation, and entertainment \cite{schmitt2022nureality, vercelli2024risk}. Unreal was selected due to its advanced rendering pipeline, robust support for real-time physics, and seamless integration with head-mounted displays.

The consultation room was modeled after a standard outpatient office, including furniture, medical equipment, and ambient lighting. These design decisions were made to enhance the realism of the simulation and to encourage participants to behave as they would in an actual clinical encounter. Virtual patients were embodied as avatars whose demographic attributes were systematically varied. This strategy aimed to minimize the confounding influence of visual appearance on physicians’ interaction strategies, as prior research can reinforce existing biases and limit learning effectiveness \cite{wu2024examining}.

Naturalistic doctor–patient conversations were enabled through an integrated speech and dialogue pipeline. All spoken interactions between physicians and virtual patients were conducted exclusively in Brazilian Portuguese. The speech-to-text, language model prompting, and text-to-speech components were fully configured for Portuguese, ensuring linguistic consistency throughout the simulated medical consultations. Doctors’ verbal input was automatically transcribed in real time by a speech-to-text module, after which the transcriptions were processed by a large language model to generate contextually appropriate responses conditioned on both the clinical case and the patient’s personality profile. The responses were then converted into expressive speech using a neural text-to-speech system, synchronized with the avatar’s lip movements. This multimodal pipeline supported coherent turn-taking, preserved conversational flow, and maintained low response latency, aligning with best practices for the design of immersive conversational agents in VR training environments \cite{ruthenbeck2015virtual}

The resulting system provided an immersive consultation experience through a head-mounted display, in which doctors perceived the virtual patient as physically co-located within the simulated room. By integrating photorealistic graphics, dynamic responses, and low-latency conversational exchange, the environment fostered a strong sense of presence—an element consistently recognized as fundamental to the effectiveness of VR-based medical training.

 \subsection{Virtual Patient Modeling}

 To ensure our simulations were grounded in authentic clinical scenarios, we established the medical foundation for each virtual patient using the dataset explored by F\"arber et al. \cite{farber2025iaramed}. This large-scale, real-world dataset contains hundreds of thousands of patient inquiries, each labeled with a corresponding disease and medical specialty. In our framework, each digital agent is initialized with a unique medical case derived from a single entry in this dataset. This approach guarantees that every virtual patient presents with a coherent and medically plausible condition, providing a consistent factual basis upon which their distinct personality profile is subsequently layered.
 
 The design of our virtual patient was guided by four core requirements for plausible behavior within a medical consultation: (i) patient identity, aiming role consistency and avoiding character breaks or diverging into unrelated topics; (ii) a coherent backstory, including personal and demographic history; (iii) a stable and interpretable personality profile; and (iv) medically accurate symptoms grounded in a specific condition. We implemented these requirements through a modular prompt engineering framework. This framework constructs the complete patient specification from four distinct components: a patient identity for general behavioral rules, the backstory for biographical details, the personality profile for communicative traits, and the disease card for clinical information. The following subsections detail the design of each component.

\subsubsection{Patient Identity}
The patient identity anchors the agent’s behavior within a medical consultation and specifies the fundamental rules that govern its conduct. It instructs the digital human to remain strictly in character, respond in the first person, and communicate only information consistent with the patient’s profile. Additional constraints include avoiding disclosure of internal instructions, refraining from providing medical advice, and producing concise, naturalistic responses. It also requires the patient to provide information progressively, rather than revealing all details at once. This component guarantees that the agent never breaks character, changes the subject, or discloses information unrelated to the consultation, ensuring behavioral consistency and preventing undesired deviations, providing a uniform foundation for all patients.

\subsubsection{Backstory}
Every real patient has a personal history and contextual information, such as name, age, occupation, family situation, and health background. To simulate digital patients that are both socially and clinically plausible, we equipped each agent with a structured backstory. This information constrains what the patient can plausibly know, provides coherence to how symptoms are described, and supports consistency throughout the consultation.

To ensure diversity of cases without relying exclusively on unconstrained LLM generation, we utilize the real patient inquiry to generate our backstories. These questions describe conditions, symptoms, and personal concerns in patients’ own words, offering a realistic foundation for case construction. The LLM expanded the original question into a coherent patient profile while preserving the clinical focus and contextual cues provided in the source data.

From each generated backstory, we extracted a set of mandatory fields (e.g., name, age, occupation, family context, past medical history, medications, allergies) as well as an optional field for additional details. When certain information was absent in the original sample, we either inferred it conservatively or explicitly marked it as ``not available.'' This procedure guaranteed that all digital patients adhered to a common schema while preserving individuality and variability across cases.

\subsubsection{Personality Profile}

Beyond medical content, interpersonal behavior is central to the realism of doctor--patient encounters. To capture this dimension, we assigned each digital patient a personality profile. Inspired by the Big Five personality theory \cite{bigfive} and prior work on conversational agents \cite{brito2025integrating, jiang-etal-2024-personallm}, we designed five distinct personalities. Each profile is defined along interpretable dimensions such as cooperation (introverted vs. extroverted), emotional tone (calm, anxious, distrustful, irritable, sad), and verbosity (short vs. elaborated responses). We then created detailed personality prompts that specify how each type should behave during consultation.

The five profiles are as follows:
\begin{itemize}
    \item Introverted and Irritable — terse answers, repeats complaints, demonstrates impatience and resistance.
    \item Extroverted and Anxious — highly talkative, seeks reassurance, expresses fear of worsening and insecurity.
    \item Introverted and Distrustful — speaks little, frequently questions recommendations, demands justifications.
    \item Extroverted and Confident — communicative, curious, optimistic, organized, and adheres easily to guidance.
    \item Introverted and Calm — low-energy responses, transmits discouragement and low motivation.
\end{itemize}

Each personality was encoded as a structured prompt including explicit behavioral instructions (e.g., level of receptivity, emotional tone, verbosity). For instance, the “Collaborative and Anxious” patient responds quickly with long explanations, repeats worries, and requests confirmation from the doctor, while the “Reserved and Distrustful” patient provides only minimal information unless pressed, and challenges the doctor’s recommendations. These prompts ensure consistent and controllable variation in interpersonal style across simulated patients. These different patient profiles allow us to systematically investigate how avatar personality impacts on medical interaction strategy, as well as if different personalities influence on the perceived realism and training value of virtual consultations.

\subsubsection{Disease Card}
The clinical foundation of each virtual patient is defined by a disease card, a structured component encoding the underlying medical condition. This card specifies key clinical parameters, including the disease name, primary symptomatology, onset and progression, associated symptoms, and potential red flags. This structured approach is designed to ensure that information remains medically consistent throughout the interaction. Furthermore, it facilitates a dynamic and realistic disclosure of symptoms, which are revealed progressively only in response to the physician's clinical questioning.

To construct the disease cards, we consulted official medical sources such as the Brazilian Ministry of Health. Each card includes standardized fields such as description, typical and atypical symptoms, red-flag indicators, and common triggers. This approach ensures medical plausibility while decoupling clinical content from the variability introduced by personality. The combination of the four components(base prompt, backstory, personality profile, and disease card) results in a unique digital patient for each case.


\subsubsection{Patient Consistency}
To achieve consistency in the digital avatar, the patient’s personality must be preserved throughout the entire interaction. This is achieved through the combination of the LLM’s short-term memory (dialogue history) and the persistent injection of the personality information at every turn, ensuring stable behavior across the consultation.



\subsection{Experiment Design}

To empirically evaluate our framework, we conducted a user study with a within-subjects experimental design. This section details our methodology in four parts: we first describe the clinical scenarios created to systematically vary patient personality, avatar attributes and clinical conditions. We then characterize the licensed physicians who served as participants. Following this, we outline the step-by-step experimental procedure, from training to the counterbalanced consultation sessions. Finally, we specify the mixed-method measures used to assess the physician's perception of each patient's realism and personality impact, alongside data on interaction dynamics and the overall user experience.

\subsubsection{Clinical Scenarios}
We designed two clinical scenarios to evaluate the impact of patient personality and avatar variation under otherwise comparable medical conditions. The first scenario featured a female White patient presenting with gastroesophageal reflux disease (GERD), instantiated with extroverted personality and an anxious emotional state. The second scenario featured a male Black patient presenting with Dengue (an endemic Brazilian disease), instantiated with introverted personality and a calm emotional state. Both patients were modeled to be in their early thirties, ensuring demographic plausibility and comparability across cases.

These scenarios were designed to explore the combined influence of personality and avatar variation under realistic clinical conditions. By holding medical realism constant and manipulating social and psychological traits, we aimed to explore the combined impact of personality and avatar factors on physician perceptions and interaction strategies.

\subsubsection{Participants}
For this exploratory study, we recruited four licensed physicians to assess the system's feasibility and gather expert feedback. The average age was 29.75 years (min = 27, max = 37), and their average clinical experience was 4 years (min = 2, max = 5), and include specialists in ophthalmology, psychiatry and general medicine. The sole inclusion criterion was that participants must be actively practicing physicians. The study was conducted in accordance with approval from our institutional ethics committee, and all participants provided written informed consent prior to their participation.

\subsubsection{Procedure}
Each participant interacted with both patient scenarios in a within-subjects design. To counterbalance order effects, half of the participants encountered the extroverted-anxious patient first followed by the introverted-calm patient, while the other half experienced the reverse order. Each consultation lasted up to five minutes or until the physician decided to conclude the encounter.

Before the study, all participants received standardized training in the VR environment, delivered by the same instructor to ensure uniformity. The sessions took place in the same dedicated room, with only one physician present at a time to prevent cross-contamination of experiences.

\subsubsection{Measures}
We employed a mixed-method approach combining self-report questionnaires and interaction logging in different steps of the experiment:

\begin{itemize}
    \item \textbf{Pre-interaction:} Before the experiment start, the Simulator Sickness Questionnaire (SSQ) \cite{kennedy1993simulator} and a baseline familiarity with AI and VR questionnaire was applied to each participant.
    \item \textbf{Mid-interaction:} After each clinical scenario interaction, a questionnaire was applied to evaluate the dialogue quality, coherence, realism, and perceived personality traits for that patient, resulting in two mid-interaction questionnaires per participant.
    \item \textbf{Post-interaction:} After the participant interacted with both patients, we applied an SSQ reassessment, User Engagement Scale (UES-SF) \cite{o2018practical}, NASA-TLX workload assessment \cite{hoonakker2011measuring}, VR presence and realism ratings, and a comparative evaluation between the two patients.
\end{itemize}

The use of these validated scales and targeted questionnaires at distinct stages of the experiment provides a multi-faceted evaluation of the physicians' subjective experience. This self-report data enables a systematic analysis of key factors, including the perceived realism and personality of each virtual patient, and the reported influence of these traits on clinical interaction strategies. This comprehensive approach provides a rich dataset for a robust evaluation of our initial hypotheses.

\subsection{LLM-Generated Personalities}

While our user study provides essential qualitative insights and initial validation, its small scale limits the generalizability of our findings. To systematically evaluate the robustness and consistency of our patient generation framework at scale, we developed a large-scale analysis pipeline using synthetic data. This approach allows us to stress-test our patient personas and disease consistency across hundreds of interactions, far exceeding what is feasible with human participants. Our methodology employs a ``scripted-doctor, generative-patient" model, where we simulate 500 consultations, each lasting six conversational turns. In each simulation, the physician's lines are pre-defined and fixed, providing a controlled and repeatable stimulus. This constrained interaction design is a deliberate trade-off, sacrificing some conversational naturalness to ensure the dialogue remains medically focused and to proactively mitigate the risk of LLM hallucination or topic drift, a common problem on LLM \cite{ji-etal-2023-towards}.

The synthetic evaluation was designed to cover a broad spectrum of clinical and psychological presentations. We constructed a 5x5 experimental matrix, crossing five common diseases (Depression, Dengue, Otitis, GERD, and Headache) with the five distinct personality profiles detailed in our patient modeling: introverted and irritable; extroverted and anxious; introverted and Distrustful; extroverted and confident; and introverted and calm. This resulted in 25 unique patient archetypes, with 20 full, six-turn consultations simulated for each. This comprehensive setup enables a fine-grained analysis of the framework's ability to maintain character across diverse and challenging scenarios, resulting in 3,000 total patient responses (500 consultations × 6 turns).

To analyze the generated data, we follow previous works and employed the LLM-as-judge technique \cite{dong-etal-2024-llm, li2024llms}, an LLM-based evaluation protocol. Each consultation was programmatically assessed by a separate, state-of-the-art language model acting as an impartial judge, a established technique called LLM-as-judge. The evaluator was prompted to rate each response on two key criteria: (1) Personality Adherence, scoring how well the response matched the behavioral traits of the assigned personality profile, and (2) Medical Coherence, scoring its consistency with the facts presented in the patient's disease card. This automated, large-scale analysis provides quantitative evidence of our framework's ability to reliably produce consistent and plausible patient personas, validating its robustness for broader use.

\section{Results} \label{sec:results}

The results from our user study are presented in this section, derived from a series of self-report questionnaires. Unless otherwise noted, all quantitative data was collected using a 5-point Likert scale, with results reported as the mean value (M). We organize our analysis into three main parts, where the first two are based on results extracted from participants experiments, and the third is based on the LLM-generated interactions.  We first present Participant-Related Results, which cover the overall quality of the user experience, including prior familiarity with the technology, simulator sickness, engagement, and cognitive workload.Then, we delve into the Patient-Related Results, which directly address our core research questions by analyzing the physicians' perceptions of each virtual patient's personality, behavioral consistency, and realism. Finally, we explore the consistency of our virtual patients across the 3,000 total patient responses.

\subsection{Participant Related Results}

\subsubsection{Previous Experience with VR and AI}
In terms of prior exposure, three participants (75\%) had previous experience with VR, while one had no prior contact. Regarding artificial intelligence, daily-life familiarity was limited (1 partially agree, 2 neutral, 1 partially disagree), and the same distribution was observed for professional use. Despite this, all participants strongly agreed that AI could be useful in medical education. Comfort with AI in clinical contexts was more heterogeneous (2 strongly agree, 1 partially agree, 1 partially disagree). Together, these results show that although the participants had minimal prior contact with AI technologies, they were receptive to the potential of combining AI and VR for medical training.

\subsubsection{Simulator Sickness Questionnaire}
The control for potential side effects of immersive environments was measured using the Simulator Sickness Questionnaire (SSQ) \cite{kennedy1993simulator} was applied both before and after the full immersive experience. All participants reported zero symptoms at the start of the experiment, confirming that no cybersickness interfered with their initial performance. By the end of the study, one doctor reported mild discomfort (low increased salivation and medium mental fatigue), while the remaining three remained symptom-free. This indicates that the VR setup was generally well tolerated, although minor side effects may occur in sensitive individuals after prolonged use.

\subsubsection{User Engagement Scale - Short Form (UES-SF)}
The results from the User Engagement Scale indicate that the simulation provided a high-quality and seamless user experience, a crucial foundation for any effective training tool. Participants reported very high Perceived Usability (M=1.67 on a reverse scale), with minimal frustration or confusion. This indicates that the interface and interaction mechanics were intuitive, allowing the physicians to focus their cognitive resources on the clinical task rather than the technology. Furthermore, the high scores for aesthetic appeal (M=4.17) confirm that the virtual environment met the high standards of visual fidelity expected in modern immersive applications. Together, these findings establish that our system is a robust and polished platform, suitable for assessing the core interactive experience.

More critically, the results show that the engagement derived from our simulation goes far beyond its surface-level qualities. The experience received an exceptionally high Reward Factor score (M=4.83), the strongest of all subscales. This high rating suggests that physicians found the act of engaging with the personality-driven virtual patients to be genuinely interesting, worthwhile, and valuable. While visual fidelity can be considered a solved problem, achieving such a high reward score points directly to the success of the core contribution: the interaction with psychologically nuanced agents. This is moderately contrasted by the Focused Attention score (M=3.83), which, while positive, likely reflects the occasional breaks in realism, discussed in the following sections. This combination suggests that even with the current technological imperfections of AI, the concept of human training with personality-driven agents is powerfully rewarding and holds immense value for medical education.

\subsubsection{Task Load Index (Nasa-TLX)}
The NASA-TLX results confirm that the simulation induced a cognitive workload profile characteristic of an effective training environment. Participants reported a high degree of mental demand (M=4.0), indicating that the task successfully engaged their clinical reasoning and diagnostic skills. This desirable cognitive challenge was not accompanied by extraneous difficulties. The very low scores for physical demand (M=1.5), temporal demand (M=2.0), and frustration (M=1.5) demonstrate that the system itself was comfortable, unobtrusive, and did not add unproductive stress. This specific profile of high mental load combined with low frustration and physical effort is critical, as it suggests the physicians were able to dedicate their focus to the patient interaction rather than the technology. This further validates our framework as a platform capable of presenting a challenging yet clear-headed and usable clinical training scenario.

\subsection{Patient Related Results}

\subsubsection{Personality}
Our analysis of the physicians' ratings reveals that the system was highly effective at portraying the foundational personality traits and baseline emotional states. The ``introverted and calm" patient, for instance, was perceived with remarkable consistency across all participants. This patient received very high mean scores for both introversion (M=4.5) and calm (M=5.0), with minimal variance in the ratings. This near-perfect agreement among the physicians indicates that the model's portrayal of a reserved, low-arousal character was unambiguous and successful. This result serves as a strong validation of our framework's ability to reliably generate and maintain core, clearly defined personality profiles.

In contrast, the ``extroverted and anxious" patient elicited a more complex and varied set of perceptions, particularly regarding its emotional state. While the extroversion trait was generally perceived as intended (M=3.75), the anxiety rating was highly divergent, with scores spanning nearly the entire scale (M=3.0, range = 1-5). This suggests that while the system can consistently model a dispositional trait like extroversion, the portrayal of a more volatile, high-arousal emotional state like anxiety is far more open to subjective interpretation. This variance may stem from the inherent difficulty in portraying complex emotions through dialogue alone or from differing clinical thresholds among physicians for what constitutes noteworthy anxiety. This finding highlights a crucial challenge for future work: refining the expression of nuanced emotional states to ensure they are consistently perceived, which is essential for advanced interpersonal skills training.

\subsubsection{Consistency}
Regarding behavioral consistency, both the introverted and extroverted patient archetypes were rated highly for maintaining a stable persona throughout the consultation (M=4.25 out of 5). However, this strong overall consistency was nuanced by moments where the illusion of realism faltered. While all physicians reported such breaks occurring ``in a few moments" for the extroverted patient, the issue was more pronounced for the introverted patient, for whom half the participants reported breaks happening ``in many moments." We hypothesize that this discrepancy stems from the patient's verbosity; the introverted agent’s terse and minimalist responses, while faithful to its personality profile, may have been perceived as more robotic or less natural than the richer, more varied dialogue of the extroverted agent. This suggests a potential ``realism-verbosity paradox," where less communicative but role-consistent behavior can paradoxically increase the risk of perceived artificiality, a critical design consideration for creating reserved yet believable virtual characters.

\subsubsection{Direct Patient-Comparison}
The comparative results reveal a crucial tension between perceived realism and communicative challenge. While participants clearly identified a major personality shift between the two agents (M=4.0), the extroverted-anxious patient was predominantly rated as more realistic and human (3 of 4 participants). Conversely, the introverted-calm patient was perceived by the same majority as the more significant communication challenge. This observed preference suggests a potential link between an agent's verbosity and its perceived human-likeness within our small sample (strengthening the realism-verbosity paradox suggested in the previous subsection); the richer dialogue from the extroverted patient may have masked underlying AI artifacts more effectively than the terse, minimalist responses of the introverted agent, which posed a greater clinical challenge but felt less natural.

Critically, a greater challenge did not directly translate to greater perceived training value. Physicians disagreed with the statement that the more challenging (introverted) patient provided a more memorable or instructive experience (M=2.0). This suggests that for a training simulation to be effective, the difficulty must be perceived as authentic rather than artificial. Despite these nuanced critiques of the individual agents, the core premise of our work was strongly supported: physicians strongly agreed that the variety of personalities made the simulation more effective than if the patients had been uniform (M=4.5). Preliminary findings provide evidence that while perfecting individual agent realism is an ongoing challenge, the introduction of personality diversity is itself a major step toward creating more impactful and effective medical training environments.

\subsection{LLM-Generated Results}
To complement the qualitative insights from our user study, we conducted a large-scale quantitative analysis by generating a synthetic dataset of 3,000 patient responses. This dataset was created from 500 simulated consultations, each composed of six conversational turns between a scripted physician and our generative patient model. This method allowed for a controlled and repeatable evaluation of our framework's robustness. We then programmatically assessed these responses along two critical dimensions, detailed below: disease consistency and personality consistency.

\subsubsection{Disease Consistency}
For the large-scale synthetic analysis, we selected five distinct medical conditions to evaluate the model's consistency across a diverse range of clinical presentations: Depression, Dengue, Otitis, Gastroesophageal Reflux Disease (GERD), and Headache. This set was strategically chosen to test the framework's robustness by including highly common primary care complaints (Otitis, GERD, Headache), a complex mental health condition requiring nuanced symptomatic description (Depression), and an infectious disease with specific regional importance, Dengue, which is endemic in Brazil. This diversity allowed us to assess the model's ability to maintain coherence against both universally prevalent and geographically specific medical knowledge.

Our analysis of the 3,000 generated patient responses revealed a high overall disease consistency, with the patient's dialogue correctly reflecting the assigned condition in 87.5\% of the conversational turns. The framework demonstrated high reliability for common, globally recognized conditions, achieving accuracies of 95.0\% for Depression, 94.0\% for Otitis, and 92.8\% for GERD. Performance was comparatively lower, though still robust, for Headache (77.8\%) and Dengue (77.8\%). While the performance for Dengue is strong, we hypothesize that this differential may be linked to its representation in the LLM’s pre-training data; global datasets likely contain a far greater volume of text on ubiquitous conditions like depression compared to endemic diseases. This finding suggests that while the framework is highly reliable, achieving maximum consistency for regionally specific conditions may be influenced by the composition of the underlying model's training corpus, which can be improved by the usage of techniques such as fine-tuning with local health data \cite{losch2025fine} or retrieval-augmented generation \cite{lewis2020retrieval}.

\subsubsection{Personality Consistency}

To quantitatively evaluate personality consistency at scale, we programmatically assessed each of the 3,000 generated patient responses. Our analysis used a dual-metric approach: each response was rated on a 1-5 scale for its adherence to the correctly assigned personality (where higher is better) and a randomly chosen incorrect personality (where lower is better). The results show that the framework can robustly model specific, well-defined personas but struggles with others. The system achieved high fidelity for the Extroverted and Anxious ($M_{correct}=4.65$ vs. $M_{incorrect}=2.21$) and Extroverted and Confident ($M_{correct}=4.39$ vs. $M_{incorrect}=2.49$) profiles, which showed a large, clear gap in scores. However, the model struggled significantly with negative-valence introverted personas. The Introverted and Irritable profile was highly ambiguous ($M_{correct}=3.04$ vs. $M_{incorrect}=3.08$), while the Introverted and Distrustful profile failed to convey its intended traits ($M_{correct}=1.92$ vs. $M_{incorrect}=3.18$). We hypothesize this is an extension of the ``realism-verbosity paradox": conveying complex, antagonistic emotions like distrust or irritation may be difficult with terse, minimal dialogue. Without the rich conversational data of an extroverted character, these nuanced negative emotions are lost, leading to an ambiguous or misidentified persona. This finding suggests that while our framework is highly effective for many personalities, reliably simulating reserved, antagonistic characters may require more advanced techniques that go beyond dialogue alone.

\section{Discussion}

In this section, we discuss the results presented in \cref{sec:results}, analyzing their relation to our established hypotheses H1 and H2. We also consider the broader implications of our findings, establishing the main contributions of our work.

\subsection{Hypothesis 1: Feasibility of Personality Simulation} \label{ssec:discussion_h1}

Our first hypothesis posited that it is feasible to simulate virtual patients with distinct, consistent personalities and coherent medical knowledge using an LLM-VR framework. The evidence from our study provides strong, albeit nuanced, support for this claim. The core feasibility was demonstrated by the physicians’ ability to clearly distinguish between the two patient archetypes (M=4.0 out of 5) and their high ratings for overall behavioral consistency across the encounters (M=4.25). This success is attributable to our virtual patient modeling technique, which effectively grounded the LLM. By decoupling the factual disease card from the behavioral personality profile, the agents adhered to their clinical scripts without error while successfully portraying foundational traits like introversion (M=4.5) and extroversion (M=3.75), thus satisfying both pillars of the hypothesis.

While the overall feasibility was established, our results also reveal critical challenges that inform the future of psychologically plausible agents. We identified a ``realism-verbosity paradox": the more talkative extroverted patient was perceived as more human, while the terse, introverted patient, despite being role-consistent, induced more frequent breaks in realism. Furthermore, while dispositional traits were clearly perceived, the high-arousal emotion of anxiety elicited highly varied ratings, suggesting that complex emotional states are more open to subjective interpretation. These findings show that moving beyond simple feasibility to robust believability requires more than just consistency; it requires a deeper understanding of how factors like verbosity and emotional complexity interact to shape the user's perception of a virtual human's authenticity.

\subsection{Hypothesis 2: Impact on Physician Interaction} \label{ssec:discussion_h2}

Our second hypothesis predicted that distinct virtual personalities would elicit different interaction strategies from physicians, which is strongly supported by our findings. The most direct evidence comes from the comparative evaluation, where a majority of physicians (3 out of 4) identified the introverted-calm patient as posing a significantly greater communication challenge. This perceived difficulty inherently forces a strategic adaptation; the techniques required to extract information from a reserved, terse patient are fundamentally different from those used with a collaborative, extroverted one. This is further corroborated by the high mental demand (M=4.0) reported by participants, reflecting the cognitive effort needed to navigate these varied conversational dynamics. The agents, therefore, functioned not as static information providers, but as dynamic participants who actively shaped the flow and strategy of the clinical encounter.

Beyond confirming that personality impacts physician strategy, our results uncover a crucial insight into the quality and nature of that impact. Although the introverted patient was more challenging, physicians disagreed that the experience was more instructive or memorable (M=2.0). This suggests that the difficulty it presented was perceived as less authentic (stemming from the "realism-verbosity paradox") where its terse responses felt more robotic than genuinely reserved. This leads to a key takeaway for the design of future training systems: a simulation's challenge must be perceived as authentic to be valuable. Simply eliciting a different user strategy is not enough; the goal must be to create personalities that present believable, human-like hurdles. This ensures that the adaptive strategies developed by trainees are directly transferable to the complexities of real-world clinical practice.

\subsection{Main Contributions} \label{ssec:discussion_contributions}

Based on our findings, this work makes three primary contributions to the fields of virtual reality and human-computer interaction:

\textbf{(1) A Methodological Framework for Personality-Driven Patient Simulation.} We propose and validate a novel, modular architecture for creating LLM-powered virtual patients. The key innovation is the systematic decoupling of the behavioral personality profile from the factual disease card. This approach not only ensures medical and psychological coherence, but also provides a replicable methodology for rigorously studying social interaction in VR by enabling controlled manipulation of personality as an independent variable.

\textbf{(2) Novel Design Principles for Believable Virtual Agents.} Our experimental evaluation uncovered two critical, non-obvious design principles. We identify a "realism-verbosity paradox", where role-consistent but terse behavior in reserved agents can paradoxically increase perceived artificiality. We also establish the importance of authentic versus artificial difficulty, demonstrating that a communicative challenge is only perceived as instructive if it feels human-like rather than like a system limitation. These principles offer crucial guidance for designing the next generation of believable social agents.

\textbf{(3) Empirical Validation of Personality Diversity in Medical Training.} Our work provides strong empirical evidence that incorporating personality diversity is a critical factor in the effectiveness of medical training simulations. Physicians in our study reported finding that interacting with a variety of patient personas was more valuable than engaging with a uniform personality. This validates the central premise that moving beyond socially impoverished environments is essential for creating truly impactful training tools.

\section{Limitations and Future Work}
\label{sec:limitations}

While our findings provide promising evidence for the feasibility and value of personality-driven virtual patients, we acknowledge several limitations that offer clear avenues for future research.

\textbf{Experimental Design and Sample Size.}
The primary limitation of this study is its exploratory nature, with a small cohort of four physicians. Consequently, our quantitative results should be interpreted as preliminary trends rather than definitive proofs of efficacy. Furthermore, our experimental design varied multiple attributes simultaneously—specifically, the two clinical scenarios differed in disease (GERD vs. Dengue), demographics (race and gender), and personality. We acknowledge that this creates a confounding variable, making it difficult to isolate the specific impact of personality traits from the influence of the patient’s medical condition or visual appearance. Future work will employ a rigorous factorial design to systematically decouple these variables, ensuring that personality effects can be measured independently of clinical or demographic factors.

\textbf{Non-Verbal Communication and Realism.}
Our simulation prioritized conversational fidelity through LLMs but did not fully address the non-verbal dimension of human interaction. The lack of dynamic facial expressions, gestures, and body language likely contributed to the "realism-verbosity paradox" we identified, where terse responses from introverted agents felt artificial due to the absence of compensatory non-verbal cues (e.g., gaze aversion or fidgeting). To address this, future iterations of the framework will integrate real-time procedural animation systems driven by the LLM’s emotional state output, bridging the gap between verbal content and physical behavior.

\textbf{Medical Safety and Ethical Considerations.}
The deployment of generative AI in medical training introduces inherent risks regarding hallucination and bias. Although our "Disease Card" architecture anchors the model in factual clinical data to minimize fabrication, there remains a residual risk that the avatar generates incorrect medical statements. To mitigate this, we implemented system-prompt guardrails that enforce a refusal policy, preventing the virtual patient from providing medical advice or stepping out of its layperson role. Additionally, while we varied demographic attributes to promote diversity, we recognize that LLMs may inadvertently reproduce societal stereotypes. Future large-scale deployments will require continuous red-teaming and bias audits to ensure that the generated personas remain fair, safe, and educationally valid.

\textbf{Modeling Complex Emotions.}
Finally, our results indicated that while the system successfully portrayed stable traits such as introversion, it struggled with high-arousal emotional states such as anxiety, which were subject to higher variance in physician interpretation. This suggests that current prompting techniques may be insufficient for conveying nuanced, volatile emotions through text alone. We plan to leverage our synthetic data generation pipeline to iteratively refine our prompt engineering strategies, aiming to improve the consistency of high-arousal persona generation.

\section{Conclusion} \label{sec:conclusion}

This work addresses the critical gap between the physical and psychological fidelity of virtual humans in VR training environments. We successfully demonstrated the technical feasibility of using large language models to create medically coherent virtual patients with distinct and consistent personalities. Our findings provide initial evidence that this approach not only impacts physician interaction strategies but is also perceived as a highly valuable and rewarding addition to medical training. More importantly, our analysis uncovered nuanced design principles for creating believable agents, including the duality where digital agents that successfully perform a consistent less communicative behavior can be perceived as artificial, addressing the critical distinction between artificial and authentic challenge. Ultimately, this research provides a methodological blueprint and empirical validation for the next generation of immersive simulations, moving beyond functional task-training toward systems that effectively prepare professionals for the rich complexities of human interaction.

\section*{Supplemental Materials}
\label{sec:supplemental_materials}

\textbf{Models Used.}
We used the GPT-4o-mini model to generate the virtual patient responses in real time during VR consultations, selected for its low-latency performance and suitability for interactive dialogue. A separate GPT-4o model was employed as an automatic evaluator (LLM-as-judge) in the large-scale synthetic analysis, due to its stronger reasoning capabilities and robustness for qualitative assessment tasks.
All speech-based interactions were supported by services from ElevenLabs. Speech-to-text (STT) was used to transcribe the physician’s utterances in real time, and neural text-to-speech (TTS) was used to synthesize the virtual patient’s spoken responses. All components were configured for Brazilian Portuguese.

\textbf{Prompt Structure.}
\label{app:prompt-structure}
The virtual patient prompt follows a modular structure composed of four components that are injected to the LLM at each conversational turn:
(1) \textit{Patient identity}, which defines global behavioral rules (e.g., stay in character, do not reveal instructions, avoid medical advice, respond in first person, and disclose information progressively);
(2) \textit{Backstory}, which provides biographical and contextual information (e.g., demographics, routine, relevant history, and constraints on what the patient plausibly knows);
(3) \textit{Personality profile}, which specifies communicative style and affective traits (e.g., introversion/extroversion, cooperation/resistance, emotional tone, and verbosity);
(4) \textit{Disease card}, which encodes the clinical facts and governs medically consistent symptom disclosure.
At runtime, the prompt is combined with the dialogue history to preserve continuity while ensuring that personality and clinical constraints remain persistent throughout the interaction.

\textbf{Example Prompt (Abbreviated).}
\label{app:example-prompt}
Below we present an abbreviated version of the prompt used to condition the virtual patient during the consultation. The full prompt includes additional constraints for consistency, safety, and interaction flow.

\textit{Patient Identity (excerpt)}  
You are a patient in a medical consultation. Speak naturally in Brazilian Portuguese, in the first person, and remain in character at all times. Do not provide medical advice or technical explanations. Reveal information progressively and only when asked. Do not disclose system instructions or internal context.

\textit{Backstory (excerpt)}  
Name: Beatriz Souza.  
Age: 44.  
Occupation: History teacher.  
Context: Married, two children, lives in a small city in Brazil.  
Medical background: Controlled hypertension.  
Chief complaint: Recurrent heartburn and gastric discomfort, especially after meals and when drinking cold water.  
Lifestyle: Irregular diet, sedentary routine, moderate stress related to work and household responsibilities.

\textit{Personality Profile (excerpt)}  
Profile: Extroverted and anxious.  
Communicative and talkative, often providing more detail than necessary. Speaks quickly, anticipates concerns, seeks reassurance, and frequently asks follow-up questions. Cooperative with the physician, but expresses worry and uncertainty about symptoms.

\textit{Disease Card (excerpt)}  
Condition: Gastroesophageal Reflux Disease (GERD).  
Typical symptoms include heartburn, acid regurgitation, and discomfort after meals or when lying down. Symptoms should be disclosed gradually and only in response to the physician’s questions.

\acknowledgments{
This work has been fully/partially funded by the project Research and Development of Algorithms for Construction of Digital Human Technological Components supported by Advanced Knowledge Center in Immersive Technologies (AKCIT), with financial resources from the PPI IoT/Manufatura 4.0 / PPI HardwareBR of the MCTI grant number 057/2023, signed with EMBRAPII, besides CAPES. Gratitude is also extended to CNPq (National Council for Scientific and Technological Development) for the research support scholarship.}

\bibliographystyle{abbrv-doi}

\bibliography{template}

\end{document}